\documentclass[sigconf, nonacm=true]{acmart} 

\usepackage{settings}
\AtBeginDocument{%
  \providecommand\BibTeX{{%
    \normalfont B\kern-0.5em{\scshape i\kern-0.25em b}\kern-0.8em\TeX}}}

\setcopyright{none}
\copyrightyear{}
\acmYear{}
\acmDOI{}

\begin{document}
\title{Low-depth Quantum Circuit Decomposition of Multi-controlled Gates}

\author{Thiago Melo D. Azevedo}
\orcid{0000-0002-1068-1618}

\author{Jefferson D. S. Silva}
\orcid{0000-0002-2063-6140}

\author{Adenilton J. da Silva}
\orcid{0000-0003-0019-7694}

\affiliation{%
  \institution{Centro de Informática \\ Universidade Federal de Pernambuco}
  \city{Recife}
  \state{PE}
  \country{Brazil}}

\begin{abstract}
Multi-controlled gates are fundamental components in the design of quantum algorithms, where efficient decompositions of these operators can enhance algorithm performance. The best asymptotic decomposition of an n-controlled X gate with one borrowed ancilla into single qubit and CNOT gates produces circuits with degree 3 polylogarithmic depth and employs a divide-and-conquer strategy. In this paper, we reduce the number of recursive calls in the divide-and-conquer algorithm and decrease the depth of n-controlled X gate decomposition to a degree of 2.799 polylogarithmic depth. With this optimized decomposition, we also reduce the depth of n-controlled SU(2) gates and approximate n-controlled U(2) gates. Decompositions described in this work achieve the lowest asymptotic depth reported in the literature. We also perform an optimization in the base of the recursive approach. Starting at 52 control qubits, the proposed n-controlled X gate with one borrowed ancilla has the shortest circuit depth in the literature. One can reproduce all the results with the freely available open-source code provided in a public repository.
\end{abstract}




\maketitle

\section{Introduction}
Quantum computing~\citep{nielsen2010quantum} is an emerging field of science and technology that leverages the quantum properties of superposition and entanglement and promises significant speedups over classical algorithms in solving problems such as physical simulations~\citep{feynman2018simulating}, unstructured search~\citep{grover1997quantum}, and number factoring~\citep{shor1997polynomial}. The current generation of devices, known as Noisy Intermediate-Scale Quantum (NISQ) systems~\citep{preskill2018quantum}, is constrained by a limited number of qubits and high noise levels. Therefore, developing more efficient quantum devices is crucial for quantum computing practical applications~\citep{kim2023evidence}.

One approach to enhance the performance of quantum devices (especially Noisy Intermediate-Scale Quantum (NISQ) devices) 
is the improvement of system software~\citep{yang2024quantum}. This pursuit has given rise to various methodologies to improve quantum computing processes. For instance, quantum state preparation~\citep{zhang2021low, zylberman2024efficient, gui2024spacetime}, quantum error mitigation strategies~\citep{cai2023quantum, gonzales2023quantum}, and decomposition of unitary gates into practical quantum circuits~\citep{barenco_1995,shende2006synthesis, iten2016quantum, adenilton2022linear, claudon2024polylogdepth}.

In the domain of quantum computing, multi-controlled operators are system software components~\citep{barenco_1995,shende2006synthesis, iten2016quantum} important for efficient quantum circuit decomposition. Their applications span various areas, including quantum neural networks~\citep{li2020qdcnn, iten2016quantum, malvetti2021quantum, shende2006synthesis}, quantum search algorithms~\citep{grover1997quantum}, quantum RAMs ~\citep{park2019circuit}, as well as state initialization protocols that establish the requisite quantum states for computation~\citep{zhang2021low, plesch2011quantum}. 

Ref.~\citep{barenco_1995} introduced several circuit decompositions for quantum gates, including a quadratic depth decomposition for general multi-controlled $U(2)$ gates without auxiliary qubits and decompositions for $SU(2)$ gates, with linear depth and without auxiliary qubits. Subsequently, Ref.~\citep{iten2016quantum} optimized several results from Ref.~\citep{barenco_1995}, particularly for Toffoli gates, by using diagonal gates to reduce the number of CNOTs needed for implementation by half. Later, Ref.~\citep{adenilton2022linear} introduced an ancilla-free decomposition for multi-controlled $U(2)$ gates with a linear depth by generalizing the previous results for $X$ gates~\citep{saeedi2013linear}.

Ref.~\citep{claudon2024polylogdepth} introduced a decomposition for $\nctrl$-controlled $X$ gates with polylogarithmic depth using one borrowed auxiliary qubit with a divide-and-conquer strategy. This decomposition technique employs a divide-and-conquer approach to create circuits with a depth of $\mathcal{O}\left(\log^{3} (\nctrl)\right)$. The application of the polylogarithmic depth multi-controlled X gates into the decomposition of multi-controlled $SU(2)$ and approximate $U(2)$ gates leads to decompositions of these gates with depths $\mathcal{O}\left(\log^{3} (\nctrl)\right)$ and $\mathcal{O}\left(\log^{3} (\nctrl) \cdot  \log (1/\epsilon) \right)$, respectively.

This work introduces an $O(\log^{\pfinal}(n))$ depth decomposition of $C^\nctrl X$ gates based on Ref.~\citep{claudon2024polylogdepth}. 
To achieve gate cancellation and reduce the number of recursive calls in the decomposition, we can invert selected multi-controlled X gates.  
In the recursive decomposition, we also empirically adjust the base case of the recursion to achieve a further lower depth. With less than 52 controls, we use the decomposition for the $n$-controlled $X$ gate described in Ref.~\citep{iten2016quantum}. Lastly, for the decomposition of $SU(2)$ or approximate gates (with error $\epsilon$), we obtain a circuit with $O(\log^{\pfinalSU}(n)$ and $O(\log^{\pfinalSU}(n)\cdot (1/\epsilon))$ depth, respectively.

The rest of this work is structured as follows. Section~\ref{sec:claudon_polylog} reviews the polylogarithmic depth decomposition for a multi-controlled $X$ gate from Ref.~\citep{claudon2024polylogdepth}. Section~\ref{sec: optimization_mcx} introduces an optimization to the polylogarithmic depth decomposition for multi-controlled $X$ gates~\citep{claudon2024polylogdepth} to reduce the order of the polylogarithmic term. Section~\ref{sec: su2-u2} applies the optimized decomposition for multi-controlled $X$ gates for multi-controlled $SU(2)$ and approximate general $U(2)$ gates. Section~\ref{sec:experiments} contains computational experiments.
\section{Polylogarithmic-depth \texorpdfstring{$C^\nctrl X$}{C n X} gates}
\label{sec:claudon_polylog}

In Ref.~\citep{claudon2024polylogdepth}, the authors introduced a recursive decomposition for $\nctrl$-controlled $X$ gates with a polylogarithmic depth that uses a borrowed auxiliary qubit. 
Let $p=\lfloor \sqrt{\nctrl} \rfloor$, the decomposition divides the $\nctrl$-qubit register $R=\{q_0, \dots, q_{\nctrl-1}\}$ into different subregisters: $R_0 = \{q_0, \dots, q_{2p-1}\}$, which has $2p$ qubits; and $b$ subregisters $R_i$, $i \in \{1, \dots, b\}$, with $p$ qubits each, except for the $R_b$, which has $r = |R \backslash R_{0}| \pmod{p}$ qubits. The $R_0$ subregister is further divided into 2 subregisters: $R_0^* = \{ q_0, \dots, q_{b-1}\}$, having the first b qubits; and $R_0^b = R_0 \backslash R_0^*$ containing the subsequent qubits. 

\begin{figure}[b]
    \centering
    \includegraphics[width=\linewidth]{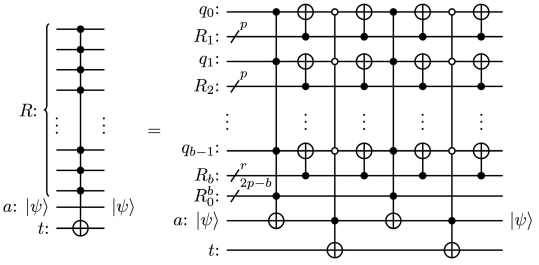}
    \caption{Decomposition of $C^\nctrl X$ gate with Polylogarithmic-depth from Ref.~\citep{claudon2024polylogdepth}. The circuit comprises the control register $R$, an auxiliary, and a target qubit. The $R$ register is divided into different subregisters:  $R_0^* = \{ q_0, \dots, q_{b-1}\}$, $R_0^b = R_0 \backslash R_0^*$ and $b$ subregisters $R_i$, $i \in \{1, \dots, b\}$ with at most $p$ qubits.}
    \Description{This image illustrates the original recursive decomposition.}
    \label{fig:claudon_dirty}
\end{figure}

The decomposition and subregisters can be seen in Fig.~\ref{fig:claudon_dirty}. The decomposition is defined as a recursion, it is applied again to the gates in the circuit with fewer controls until it reaches the base case. For the decomposition, it is ensured that there are more available borrowed qubits in register $R_0^b$ than operations to be performed in parallel so they can be decomposed recursively. 

With this strategy, a $C^{\nctrl} X$ gate can be decomposed in terms of $C^{2p}X$, $C^{b+1}X$ and $C^{p} X$ gates. Let $D_\nctrl$ be the depth of a $C^\nctrl X$ gate, $D_\nctrl$ can be written recursively in terms of $D_{2p}$, $D_{b+1}$ and $D_{p}$. Noticing that the multiple $C^{p} X$ gates act in parallel, their depth is equivalent to the depth of a single $C^{p} X$ gate, and considering the two $X$ gates for the open controls of the $C^{b+1}X$, we obtain that

\begin{equation}
    D_\nctrl = 2 D_{2p} + 4D_p + 2D_{b+1}+4.
\end{equation}

To study the asymptotic behavior of the depth, the authors make a change of variable. Since $p-2 \leq b \leq p$, for the upper bound, $\fdepth{k} \equiv  D_{2^{k+2}}$ is equivalent to the circuit depth of a $C^{{2^{k+2}}}$ gate ~\citep{claudon2024polylogdepth}. Then, we can write: 

\begin{equation}
    \fdepth{k} \leq 8 \fdepth{k/2} + 4.
\end{equation}

\noindent An asymptotic bound for recursions of the form

\begin{equation*}
    T(\Tilde{n})\leq aT(\Tilde{n}/b)+f(\Tilde{n}),
\end{equation*}

\noindent can be found by using the Master Theorem~\citep{CormenAlgorithms4th}. Since in this case, $f(k)=4$, therefore $\fdepth{k} = \mathcal{O}\left(k^3\right)$. In terms of $\nctrl$, the depth is $\mathcal{O}\left(\log^3(\nctrl)\right)$~\citep{claudon2024polylogdepth}.
\section{Optimizations for recursive \texorpdfstring{$C^\nctrl X$}{C n X} gates}
\label{sec: optimization_mcx}

\begin{figure*}[ht]
    \centering
    \includegraphics[width=\linewidth]{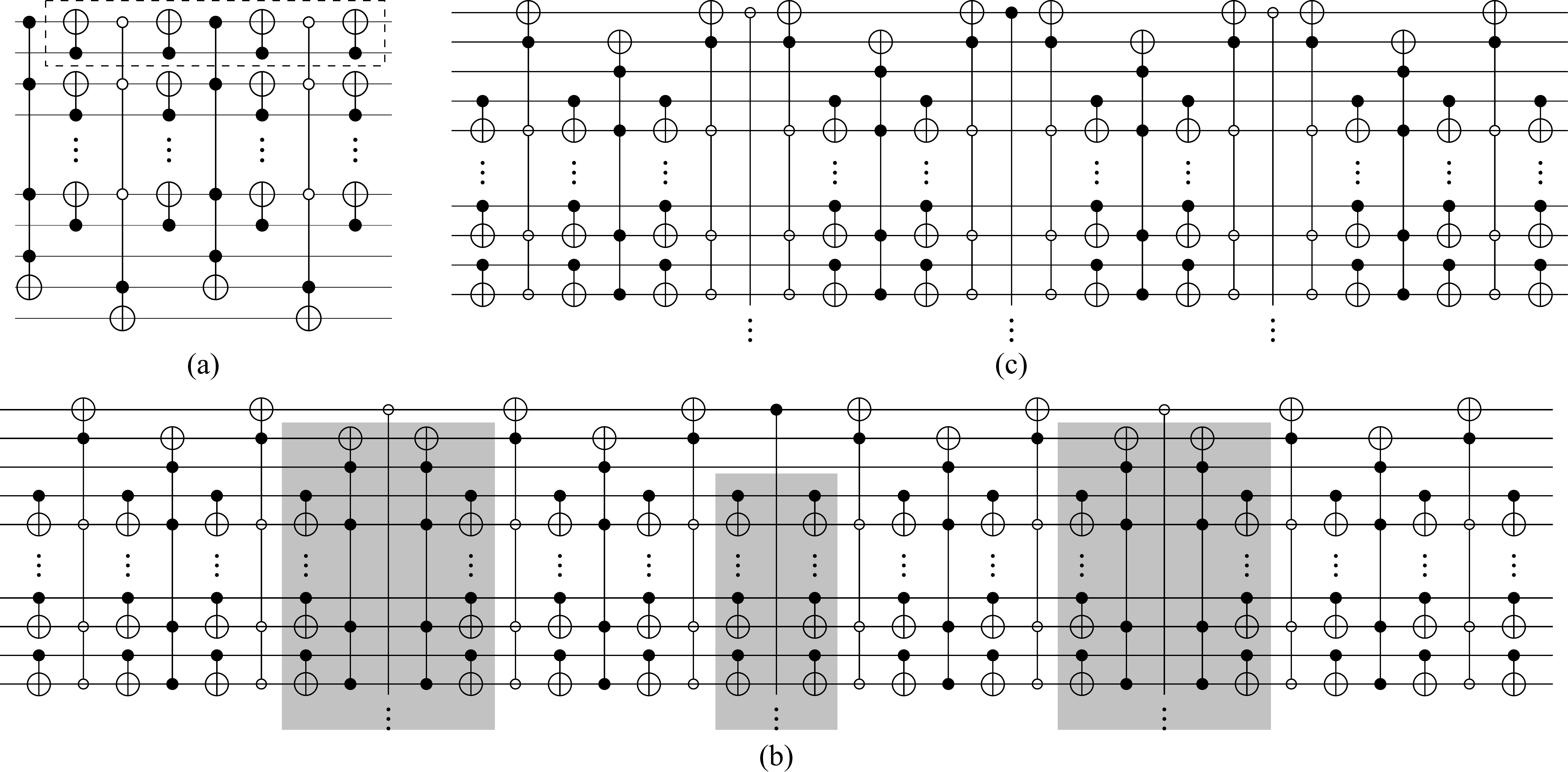}
    \caption{Gate cancelation is achieved by reversing the circuit of the first and third columns of $C^p X$ gates. We analyze the optimization for one line of gates, equivalent to the other lines. After applying the recursion again, the reversal allows the cancelation of two columns of multi-controlled gates in the first and second blocks and the third and fourth blocks. Fig. (a) represents the original recursion in which the gates highlighted in a dashed-dotted line are decomposed in Fig. (b). The last column of the second block and the first column of the third block are also canceled. Rectangles with opaque backgrounds have been added to highlight which gates are undergoing cancelation, Fig. (b). After the cancelations, the resulting circuit for each line of the original $C^p X$ gates has reduced depth and gate counts, as shown in Fig. (c).}
    \Description{This image illustrates the first optimization level.}
    \label{fig:claudon_optim_1}
\end{figure*}

The work of Ref.~\citep{claudon2024polylogdepth} can be improved by using that 
$X^{-1} = X$, so the circuit of a multi-controlled $X$ gate can be inverted without changing its result. This trick can be used in the decomposition of $C^\nctrl X$ gates from Fig.~\ref{fig:claudon_optim_1}(a), we can reverse the circuits of the first and third columns of the $C^p X$ gates completely contained in the dotted box, to achieve gate cancelations with the second and fourth columns.

Fig.~\ref{fig:claudon_optim_1} (b) is an enlargement of the dotted box in Fig.~\ref{fig:claudon_optim_1} (a), where the $C^pX$ contained in the box are replaced by a recursive application of the decomposition. In Fig \ref{fig:claudon_optim_1} (b), the gray boxes indicate gate cancelation of two columns of multi-controlled gates in the first and second blocks and the third and fourth blocks. Additionally, one column in the second and third blocks is also canceled. This leads to the $4$ blocks having a depth of $6\fdepth{k/4}$, $5\fdepth{k/4}$, $5\fdepth{k/4}$, $6\fdepth{k/4}$, respectively, as can be seen in the resulting circuit of Fig.~\ref{fig:claudon_optim_1} (c). Taking into account all the open controls, we can write an optimized recurrence relation for the circuit depth $\fdepth{k}$,

\begin{equation}
    \fdepth{k} \leq 4\fdepth{k/2} + 22\fdepth{k/4}+\mathcal{O}( 1 ).
\end{equation}

For recursive relations of the form

\begin{equation*}
    T(\Tilde{n}) \leq f(\Tilde{n}) + \sum_i a_i T(\Tilde{n}/{b_i}),  
\end{equation*}

\noindent we can use the Akra-Bazzi Theorem~\citep{AkraBazzi1998, CormenAlgorithms4th} to find asymptotical bounds for $T(\Tilde{n})$. In our case, $f(k) =\mathcal{O}( 1 )$ and we obtain an asymptotic bound of $\fdepth{k} = \mathcal{O}(k^\alpha)$, where $\alpha$ is the solution to the equation

\begin{equation}
    \frac{4}{2^\alpha}+\frac{22}{4^\alpha}=1.
\end{equation}

\noindent Solving it numerically, we can find that $\alpha=2.828$. Therefore, $\fdepth{k} = \mathcal{O} (k ^{2.828})$. Since $n = 2^{k+2}$, then $D(\nctrl)=$ $\mathcal{O}(\log^{2.828} \nctrl)$.  
The optimization can then be applied to the next level of the recursion. 

\begin{figure*}[ht]
    \centering
    \includegraphics[width=\linewidth]{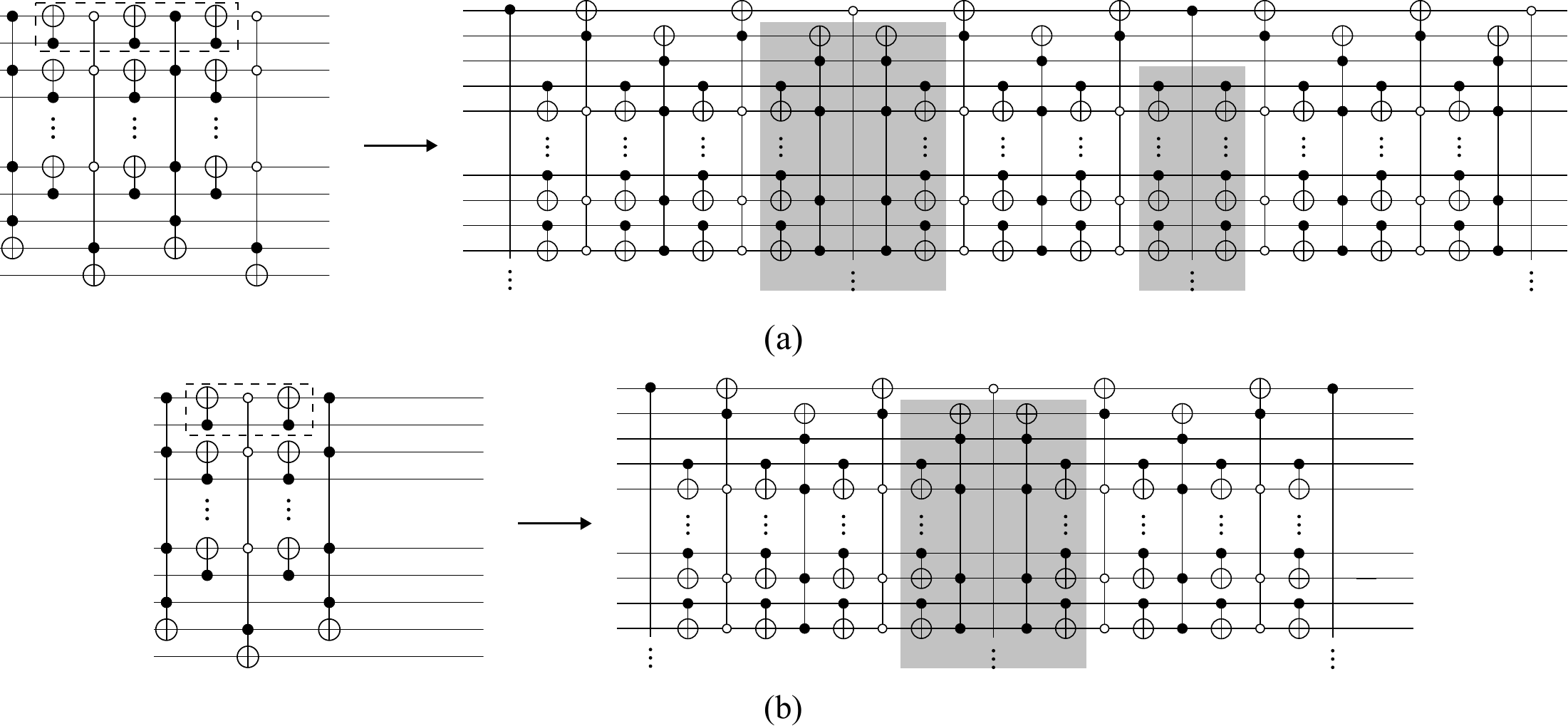}
    \caption{ Examples of circuits that can also be optimized. (a) If there are three columns of parallel $C^p X$ gates, each column is decomposed into blocks with a depth of $6\fdepth{k/2^{j+1}}$, $5\fdepth{k/2^{j+1}}$ and $7\fdepth{k/2^{j+1}}$, respectively. (b) If there are two columns of parallel $C^p X$ gates, they are decomposed into two blocks with a depth of $6\fdepth{k/2^{j+1}}$. In both cases, the same cancelations are also achieved if the multi-controlled operators acting on the ancilla or the target at the beginning or end of the circuit are absent or if the circuit is reversed.}
    \Description{This image illustrates the second optimization level.}
    \label{fig:optim_rules}
\end{figure*}

In the next recursive call, the number of adjacent $C^pX$ will be 2 or 3 as in Fig.~\ref{fig:claudon_optim_1} (c). Fig.~\ref{fig:optim_rules} shows optimizations that can be done depending on the number of columns of $C^p X$ gates in the same block. 
We consider the cases of $2$ or $3$ columns of $C^p X$ gates and the same cancelations are obtained if the block is reversed or if one of the $C^{2p}$ or $C^{b+1}$ gates on the edges of the circuit have been canceled already. When there are $3$ columns, each gate will be decomposed in a block with a depth of $6\fdepth{k/8}$, $5\fdepth{k/8}$, $7\fdepth{k/8}$, respectively, this can be seen in Fig.~\ref{fig:optim_rules} (a).  When there are only $2$ columns together in a block, which can be seen in Fig.~\ref{fig:optim_rules} (b), we obtain a depth of $6\fdepth{k/8}$ for each of the columns. 
    
From Fig.~\ref{fig:claudon_optim_1} (c), after the first round of optimizations, the circuit is divided into blocks with $3$ columns of $C^p X$ gates or $2$ columns of $C^p X$ gates. Making use of the optimizations shown in Fig.~\ref{fig:optim_rules}, the two blocks in the first half of the circuit from Fig.~\ref{fig:claudon_optim_1} (c) will have a combined depth of $6\fdepth{k/4}+30\fdepth{k/8}$. We then consider both halves and add the $4$ multi-controlled gates with depth $\fdepth{k/2}$ to obtain the following recurrence relation:

\begin{equation}
    \fdepth{k}\leq 4\fdepth{k/2}+12\fdepth{k/4}+60\fdepth{k/8}+\mathcal{O}( 1 ).
\end{equation}

By the Akra-Bazzi method~\citep{AkraBazzi1998, CormenAlgorithms4th}, $\fdepth{k} = \mathcal{O}\left(k^\alpha\right)$, where $\alpha$ is the solution to the Equation~\ref{eq:log277}.
\begin{equation}
\label{eq:log277}
    \frac{4}{2^\alpha}+\frac{12}{4^\alpha}+\frac{60}{8^\alpha}=1.
\end{equation}
Therefore, $\alpha=\pfinal$. Since $\nctrl = 2^{k+2}$, this means the depth is $\mathcal{O}\left(\log^{\pfinal} (\nctrl) \right)$. So, the optimizations in the second recursive call further reduced the depth.

\begin{theorem}\label{cnx}
    A decomposition of a $C^\nctrl X$ gate can be constructed as a circuit with asymptotic depth $\mathcal{O}\left( \log^{\pfinal} (\nctrl) \right)$.
\end{theorem}
    
\section{Optimizations for multi-controlled U(2) and SU(2) gates}
\label{sec: su2-u2}

\subsection{Multi-controlled SU(2) gates}

\begin{figure}[htb]
    \centering
    \includegraphics[width=\linewidth]{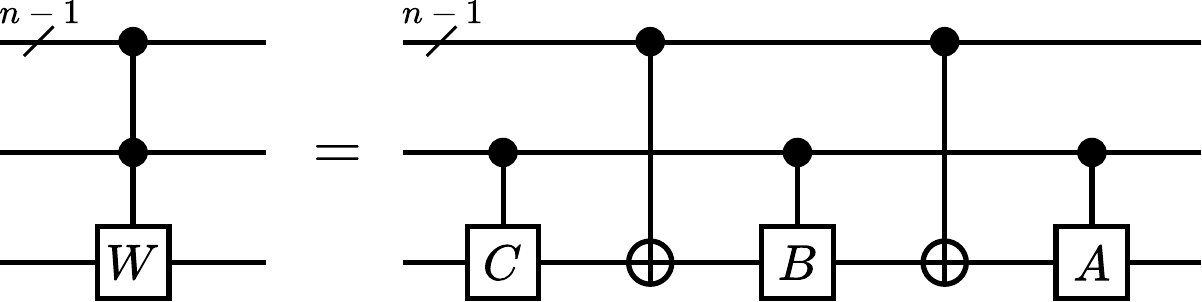}
    \caption{Decomposition of an $\nctrl$-controlled $W$ gate, where $W, A, B, C \in SU(2)$, as described in Ref.~\citep[Lemma 7.9]{barenco_1995}.}
    \Description{This image illustrates the decomposition of a multi-controlled SU(2) gate.}
    \label{fig:ncontrolled-w}
\end{figure}

We apply the optimizations developed for multi-controlled $X$ gates with an auxiliary qubit to the decomposition of multi-controlled $SU(2)$ gates. We use the multi-controlled SU(2) gate decomposition from Ref.~\citep{barenco_1995}, shown in Fig.~\ref{fig:ncontrolled-w}, and our optimized method for decomposing multi-controlled $X$ gates with one auxiliary qubit, which in this case is the $\nctrl$-th control qubit. We conclude that

\begin{theorem}\label{lem:su2}
    A decomposition of a $C^\nctrl W$ gate, $W\in SU(2)$, can be constructed as a circuit with an asymptotic depth of $\mathcal{O}\left( \log^{\pfinal} (\nctrl) \right)$.
\end{theorem}

\subsection{Approximate Multi-controlled U(2) gates}

\begin{figure}[htb]
    \centering
    \includegraphics[width=\linewidth]{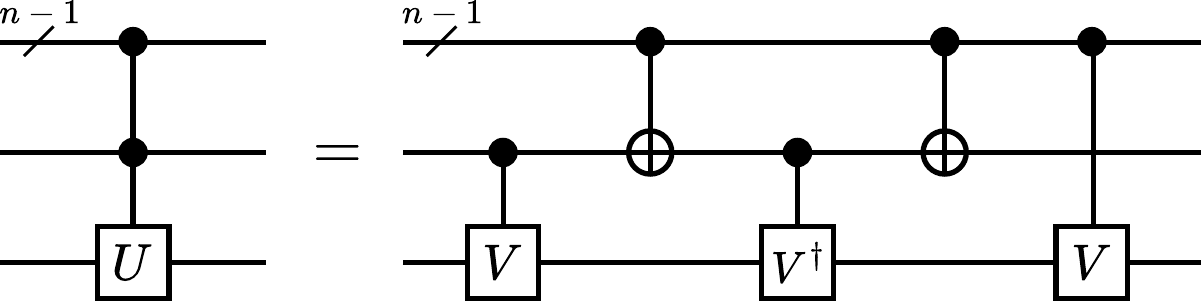}
    \caption{Decomposition of an $\nctrl$-controlled $U \in U(2)$ gate, where $U = V^2$, as described in Ref.~\citep[Lemma 7.5]{barenco_1995}.}
    \Description{This image illustrates the decomposition of a multi-controlled U(2) gate.}
    \label{fig:ncontrolled-u2-barenco}
\end{figure}

We also apply our optimized decomposition for multi-controlled $X$ gates to the approximate decomposition for multi-controlled $U(2)$ gates from Ref.~\citep{barenco_1995}, improving the result of Ref.~\citep{claudon2024polylogdepth}. The exact decomposition for a $C^{\nctrl}U$ gate, $U\in U(2)$, from Ref.~\citep{barenco_1995} is shown in Fig.~\ref{fig:ncontrolled-u2-barenco}. The decomposition is defined recursively, as the same decomposition is applied on a $C^{\nctrl-1}V$ gate, where $V^2 = U$. 

 An approximate version of this gate can be developed by halting the recursion after $\mathcal{O}(\log (1/\epsilon))$ steps~\citep{barenco_1995}. 
 
Making use of our decomposition of multi-controlled $X$ gates, in each step we achieve an asymptotic depth of $\mathcal{O}(\log ^{\pfinal})$. Considering all the $\mathcal{O}(\log (1/\epsilon))$ steps, we conclude

\begin{theorem}\label{lem:approx_u2}
    A decomposition of an approximate $C^\nctrl U$ gate, $U\in U(2)$, can be constructed as a circuit with an asymptotic depth of $\mathcal{O}\left(log^{\pfinal}(\nctrl)\, log(1/\epsilon)\right)$.
\end{theorem}

\section{Experiments}
\label{sec:experiments}

In this section, we conduct computational experiments to compare the circuit depth and number of CNOTs between our optimized decomposition of a $C^\nctrl X$ gate with an auxiliary qubit, as described in Theorem \ref{cnx}, and existing methods. We contrast our approach with the linear decomposition outlined in Ref.\citep[Lemma 9]{iten2016quantum} and the polylogarithmic recursive decomposition detailed in Ref.\citep[Proposition 1]{claudon2024polylogdepth}. We implemented the linear decomposition using the method from \citep{qclib} and the recursive decomposition as per \citep{claudon2024github}. We employed two Python libraries for our experiments, Qiskit (version 1.0.2) and qclib (version 0.1.12). Our code is available at \citep{silva2024github}.

\begin{figure}[ht]
    \centering
    \includegraphics[width=\linewidth]{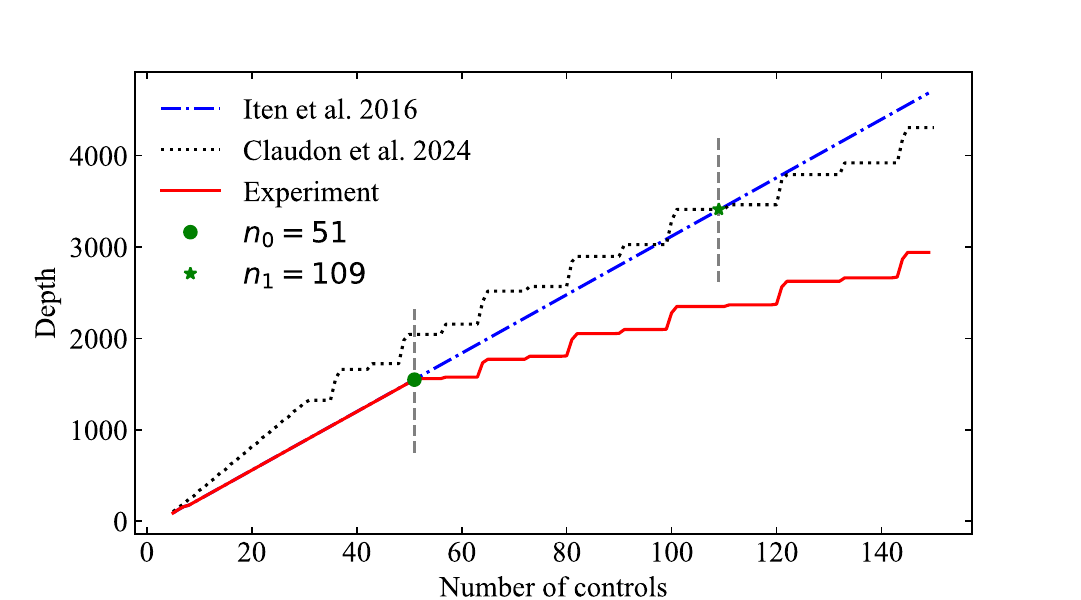}
    \caption{Comparison of circuit depth for various decompositions of multi-controlled $C^n X$ gates using an auxiliary qubit. The blue dash-dotted line illustrates the linear decomposition as described in Ref.\citep[Lemma 9]{iten2016quantum}; the black dotted line illustrates the original recursive decomposition from Ref.\citep[Proposition 1]{claudon2024polylogdepth}; and the red solid line illustrates our optimized decomposition, as detailed in Theorem~\ref{cnx}. The intersection points $n_0$ and $n_1$ mark where our optimized decomposition converges with the linear and original recursive decompositions.}
    \Description{This image illustrates the depth performance of our optimized decomposition}
    \label{fig:graph_depth1}
\end{figure}

In our implementation, we also optimized the base case of our recursion. The base case is applied when $\nctrl \leq n_b$, and we chose a value for $n_b$ that optimally reduces the depth of our decomposition. To determine the value of $n_b$, we first analyzed which choice would reduce the value of $n_0$, the final intersection between the depth of our implementation and the depth of the linear decomposition described in Ref.~\citep[Lemma 9]{iten2016quantum}. We searched for values of $n_b$ in the interval of $[4, 53]$, and obtained multiple optimal choices. We then analyzed which of the obtained values would lead to a lower depth in some sample points between $\nctrl=100$ and $\nctrl = 7000$. Ultimately, we chose $\nctrl=26$ as it achieved an overall lower depth. Additionally, in our implementation, for $\nctrl \leq 51$, we apply the linear decomposition~\citep[Lemma 9]{iten2016quantum}. For $\nctrl > 51$, we apply our optimized recursive decomposition with base case $\nctrl\leq 26$.

In Fig. \ref{fig:graph_depth1}, we compare our implementation with the linear decomposition from Ref.~\citep[Lemma 9]{iten2016quantum} and with the implementation of the decomposition from Ref.~\citep[Proposition 1]{claudon2024polylogdepth}. In the implementation of Ref.~\citep[Proposition 1]{claudon2024polylogdepth} found in \citep{claudon2024github}, the authors have set the base case of their recursion to $\nctrl \leq 30$, which we left unaltered. Fig.~\ref{fig:graph_depth1} shows that our decomposition has a lower depth than the linear decomposition from \citep[Lemma 9]{iten2016quantum} for $\nctrl > 51$. In contrast, the decomposition from \citep[Proposition 1]{claudon2024polylogdepth} only achieves this feat for $\nctrl>109$. Furthermore, we observe that the depth of our decomposition is lower than the depth from Ref.~\citep[Proposition 1]{claudon2024polylogdepth} for all evaluated values of $\nctrl$.

\begin{figure}[ht]
    \centering
    \includegraphics[width=\linewidth]{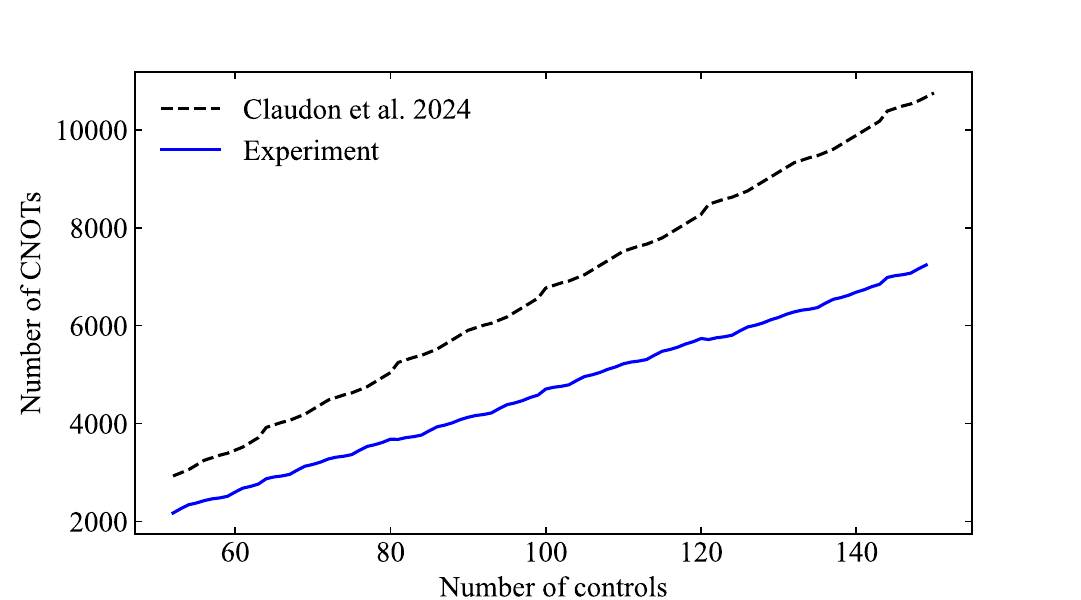}
    \caption{Comparison of the number of CNOTs, for different decompositions of multi-controlled $C^\nctrl X$ gates with an auxiliary qubit. The black dashed line illustrates the recursive decomposition described in Ref.~\citep[Proposition 1]{claudon2024polylogdepth}, and the blue solid line illustrates our optimized decomposition.}
    \Description{This image illustrates the CNOT count performance of our optimized decomposition.}
    \label{fig:graph_size}
\end{figure}

\begin{figure}[ht]
    \centering
    \includegraphics[width=\linewidth]{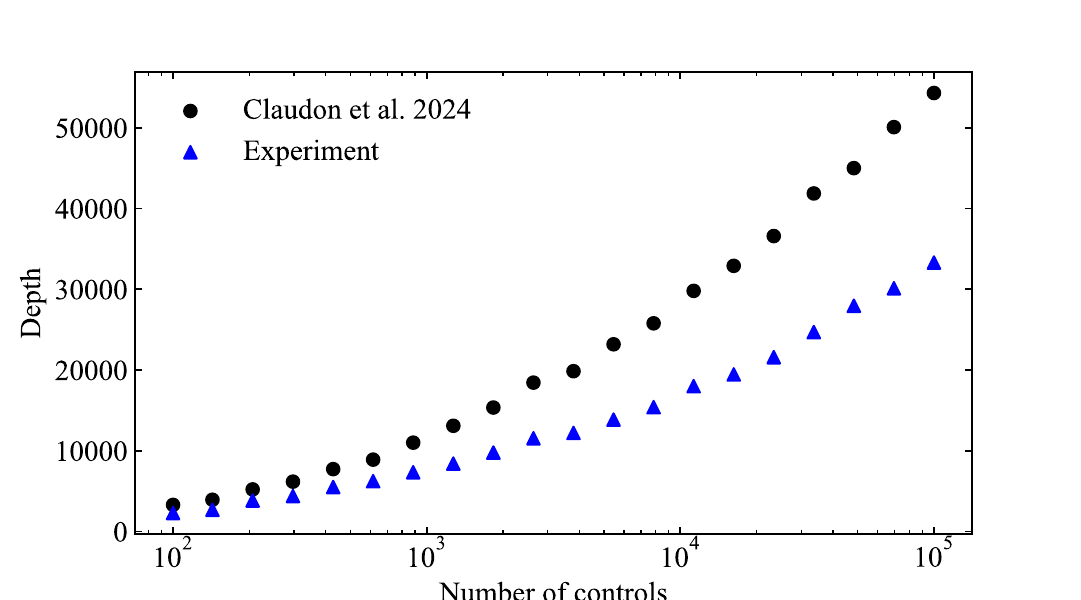}
    \caption{Comparison of the depth of different decompositions of multi-controlled $C^n X$ gates with an auxiliary qubit for larger values of $\nctrl$ on a logarithmic scale. The blue triangle markers represent the depth of the optimized recursive decomposition described by Theorem~\ref{cnx}, and the black circle markers represent the depth of the original recursive decomposition, as described in Ref.~\citep[Proposition 1]{claudon2024polylogdepth}.}
    \Description{This image illustrates the depth performance of our optimized decomposition on a logarithmic scale.}
    \label{fig:graph_depth2}
\end{figure}

Fig.~\ref{fig:graph_size} compares the number of CNOT gates of our implementation with the decomposition from Ref.~\citep[Proposition 1]{claudon2024polylogdepth} and shows that our method also significantly reduces the number of controlled operators. This improvement occurs because our optimization cancels several multi-controlled layers from the original circuit described in Ref.~\citep{claudon2024polylogdepth}, thereby reducing the number of operators in the circuit. Fig.~\ref{fig:graph_depth2} compares the circuit depth of our implementation with the decomposition from \citep[Proposition 1]{claudon2024polylogdepth} for larger values of $\nctrl$ and more evidently shows the asymptotic advantage of our optimized decomposition compared to the decomposition from Ref.~\citep{claudon2024polylogdepth}.

\section{Conclusion}

We introduced a low-depth decomposition for $\nctrl$-controlled $X$ gates with auxiliary qubits based on the decomposition from Ref.~\citep{claudon2024polylogdepth}. By selectively inverting $k$-controlled $X$ gates from the original circuit, we reduced the upper bound of the circuit depth to $\mathcal{O}\left(\log ^{\pfinal} (\nctrl) \right)$, achieving the lowest reported asymptotic circuit depth in the literature.

We apply the developed decomposition for $\nctrl$-controlled $X$ gates with auxiliary qubits to the decomposition of $\nctrl$-controlled $SU(2)$ and approximate $U(2)$ gates from Ref.\citep{barenco_1995}, achieving lower asymptotic upper bounds of circuit depth than those reported in Ref.\citep{claudon2024polylogdepth} and the lowest in existing literature. We obtained theoretical asymptotic depths of $\mathcal{O}\left(\log^{\pfinalSU} (\nctrl) \right)$ and $\mathcal{O}\left(\log^{\pfinalSU} (\nctrl) (1/\epsilon) \right)$ for $\nctrl$-controlled $SU(2)$ and approximate $U(2)$ gates, respectively.

Additionally, we implemented another optimization by empirically adjusting the base case of the recursion, which resulted in a lower overall depth. Our analysis determined that $\nctrl=26$ was the optimal value for the base case. We conducted experiments to compare our decomposition against that from Ref.~\citep{claudon2024polylogdepth}, finding that our approach achieved lower circuit depth and fewer CNOTs across all tested values of $\nctrl$.

For future work, we plan to explore further optimizations for the recursive decomposition and the potential use of additional auxiliary qubits to achieve even lower depths. Another possible research direction is investigating applications of the methods developed in this work to other circuit decompositions or different areas of quantum computing.

\section*{Data availability}
All data and software generated during the current study are available at the following sites: \url{https://github.com/qclib/qclib-papers} and \url{https://github.com/qclib/qclib}~\citep{qclib}.

\section*{ACKNOWLEDGMENTS}
This work is supported by the Brazilian research agencies Conselho Nacional de Desenvolvimento Científico e Tecnológico–CNPq (Grant No. 308730/2018-6), Coordenação de Aperfeiçoamento de Pessoal de Nível Superior (CAPES)–Finance Code 001, and Fundação de Amparo à Ciência e Tecnologia do Estado de Pernambuco–FACEPE
(Grant No. APQ-1229-1.03/21).

\bibliographystyle{ACM-Reference-Format}
\bibliography{refs}


\end{document}